\documentclass{jfm}

\usepackage{graphicx}
\usepackage{epstopdf}
\usepackage{natbib}
\usepackage{amsmath}
\usepackage{amssymb}
\usepackage{mathtools}
\usepackage{bm}
\usepackage{mathrsfs}
\usepackage{colortbl}
\usepackage[svgnames,x11names]{xcolor}
\usepackage[font=small,labelfont=bf]{caption}
\usepackage{multirow}
\usepackage{booktabs}
\usepackage{bibentry}
\nobibliography*

\ifCUPmtlplainloaded \else
  \checkfont{eurm10}
  \iffontfound
    \IfFileExists{upmath.sty}
      {\typeout{^^JFound AMS Euler Roman fonts on the system,
                   using the 'upmath' package.^^J}%
       \usepackage{upmath}}
      {\typeout{^^JFound AMS Euler Roman fonts on the system, but you
                   dont seem to have the}%
       \typeout{'upmath' package installed. JFM.cls can take advantage
                 of these fonts,^^Jif you use 'upmath' package.^^J}%
      }
  \else
  \fi
\fi

\ifCUPmtlplainloaded \else
  \checkfont{msam10}
  \iffontfound
    \IfFileExists{amssymb.sty}
      {\typeout{^^JFound AMS Symbol fonts on the system, using the
                'amssymb' package.^^J}%
       \usepackage{amssymb}%
         
         \let\geq=\geqslant
      }{}
  \fi
\fi

\ifCUPmtlplainloaded \else
  \IfFileExists{amsbsy.sty}
    {\typeout{^^JFound the 'amsbsy' package on the system, using it.^^J}%
     \usepackage{amsbsy}}
    {}
\fi

\newcommand\Rey{\mbox{\textit{Re}}}  

\newsavebox{\astrutbox}
\sbox{\astrutbox}{\rule[-5pt]{0pt}{20pt}}

\newcommand{\vol}{\mathop{\ooalign{\hfil$V$\hfil\cr\kern0.08em--\hfil\cr}}\nolimits}

\newcommand\Wo{\mbox{\textit{Wo}}}
\newcommand\A{\mbox{\textit{A}}}
\title[Non-modal transient growth of disturbances in pulsatile and oscillatory pipe flow]{Non-modal transient growth of disturbances in pulsatile and oscillatory pipe flow}

\author[Duo Xu, Baofang Song and Marc Avila]%
{Duo Xu$^{1}$\thanks{{Email address for correspondence: duo.xu@zarm.uni-bremen.de}},\ns
Baofang Song$^{2}$ 
and
Marc Avila$^{1}$}

\affiliation{
$^1$ University of Bremen, Center of Applied Space Technology and Microgravity  (ZARM), 28359 Bremen, Germany\\[\affilskip]
$^2$ Tianjin University, Center for Applied Mathematics, Tianjin 300072, China\\[\affilskip]
}

\pubyear{?}
\volume{?}
\pagerange{?}
\date{?; revised ?; accepted ?. - To be entered by editorial office}
\begin{document}

\maketitle

\begin{abstract}
Laminar flows through pipes driven at steady, pulsatile or oscillatory rates undergo a sub-critical transition to turbulence. We carry out an extensive linear non-modal stability analysis of these flows and show that for sufficiently high pulsation amplitudes the stream-wise vortices of the classic lift-up mechanism are outperformed by helical disturbances exhibiting an Orr-like mechanism. In oscillatory flow, the energy amplification depends solely on the Reynolds number based on the Stokes-layer thickness and for sufficiently high oscillation frequency and Reynolds number, axisymmetric disturbances dominate. In the high-frequency limit, these axisymmetric disturbances are exactly similar to those recently identified by \citet{Biau16} for oscillatory flow over a flat plate.  In all regimes of pulsatile and oscillatory pipe flow, the optimal helical and axisymmetric disturbances are triggered in the deceleration phase and reach their peaks in typically less than a period. Their maximum energy gain scales exponentially with Reynolds number of the oscillatory flow component. Our numerical computations unveil a  plausible mechanism for the turbulence observed experimentally in pulsatile and oscillatory pipe flow.
\end{abstract}

\begin{keywords}
\end{keywords}

\section{Introduction}\label{sec:introduction}

Physiological flows are unsteady in nature and are characterized by complex geometries and fluid-structure interaction. 
In healthy individuals, arterial flow is generally assumed laminar, but complex (\emph{disturbed}) flow patterns are acknowledged to play an important mechanistic role in the development of vascular diseases {\citep{Ku1997,Chiu2011}}. Even for the simple case of pulsatile flow in a straight pipe, the mechanisms of instability and transition to turbulence are poorly understood and particularly the dependence on the pulsation amplitude ($\A=U_o/U_s$, where $U_o$ and $U_s$ are the magnitude of the oscillatory and steady components of the velocity) is largely unknown. This makes it difficult to assess whether disturbed flow patterns in arterial and respiratory flow are solely due to geometric and structural effects (e.g.\ vessel curvature and flexibility, bifurcations), or are also related to the stability of pulsatile pipe flow. 
 
Pulsatile flow in a straight pipe is governed by the pulsation amplitude $\A$, the {Womersley} number $\Wo=D/2\sqrt{\omega/\nu}$ and the Reynolds number $\Rey_s=U_s D/\nu$. Here $D$ the pipe diameter, $\omega$ the angular frequency of the pulsation and  $\nu$ the kinematic viscosity of the fluid. The limit of small pulsation amplitude $\A\rightarrow 0$ (steady flow), is relevant to laminar blood flow in capillaries, whereas the opposite limit $\A\rightarrow\infty$ (oscillatory flow) is relevant to respiratory flow.  In humans, the airflow may be laminar, transitional or turbulent depending on the airway segment \citep{kleinstreuer2010}. The intermediate regime, in which the pulsatile flow component is similar to the steady one ($\A\gtrsim 1$), is typical of blood flow in the large arteries. 

Steady laminar pipe flow is linearly stable and  transition can only be triggered with finite-amplitude disturbances \citep{reynolds83}.  Following transition, turbulence persists provided that $\Rey_s\gtrsim2040$ \citep{Avila11}. Despite the non-linear nature of the transition, the key underlying mechanism is linear \citep{SchmidHenningson_springer2001}. The Navier--Stokes equations linearized about the laminar flow are non-normal and disturbances can be transiently amplified before asymptotically decaying. \citet{schmid1994} showed that in pipe flow the optimal (non-modal) disturbance consists of a pair of stream-wise rolls, which generate a pair of stream-wise velocity streaks (lift-up mechanism). The perturbation's energy gain in this process scales as $G\propto Re_s^2$, and keeping the flow laminar as $Re_s$ increases becomes an arduous task. 

Oscillatory pipe flow is linearly (Floquet) unstable when the Reynolds number based on the Stokes-layer thickness, $\Rey_\delta=U_o\delta/\nu\gtrsim 10^3$, where  $\delta = \sqrt{2\nu/\omega} $ is the thickness of the Stokes layer \citep{Thomas12}. In experiments turbulence was observed already for $280 \lesssim \Rey_\delta \lesssim 550$ \citep{Sergeev66,Merkli75,Hino76,Eckmann91,zhao1996}, indicating that oscillatory pipe flow also undergoes transition via finite-amplitude disturbances. {\citet{Feldmann12,Feldmann16} performed direct numerical simulations of oscillatory pipe flow initialized with fully turbulent fields and confirmed the existence of sustained turbulence in the sub-critical regime.} \citet{Thomas11} showed that the linear instability of oscillatory pipe flow persists also for pulsatile flow, however in experiments transition occurs much earlier \citep[see e.g.][]{Sarpkaya66, Stettler86,Xu17,xu2020}. Taken together these results suggest that pulsatile pipe flow undergoes a sub-critical transition to turbulence in all regimes.

\citet{xu2020} recently reported on a nonlinear instability of pulsatile pipe flow, which occurs at pulsation amplitudes relevant for arterial flow. In their experiments, geometric imperfections triggered a helical wave pattern which emerged cyclically during the deceleration phase and broke down into turbulence, before decaying.  For $\A=1$ they observed transition at Reynolds numbers as low as $\Rey_s\approx 800$. In addition, \citet{xu2020} performed also linear non-modal transient growth computations at a selected parameter set and showed that the most amplified disturbance is a helical wave. Direct numerical simulations initialized with this helical wave reproduced the flow patterns and the time of turbulence breakdown observed experimentally, which suggests an important role of transient growth. {\citet{Nebauer2019} anticipated that in certain regimes of pulsatile pipe flow, the energy growth scales exponentially with the Reynolds number. This was also reported for the oscillatory Stokes layer over a flat plate \citep{Biau16} and for the flow following an axisymmetric stenosis \citep{Blackburn08JFM}.  Large non-modal transient amplification of disturbances has also been found in pulsatile channel flow \citep{Tsigklifis17}.} In what follows, we present a comprehensive non-modal linear analysis of the sub-critical regimes of pulsatile and oscillatory pipe flows.

\section{Methods}\label{sec:methods}

We consider an incompressible viscous fluid driven at a pulsatile flow rate in a straight pipe of circular cross-section. Lengths, velocities and time are rendered dimensionless with $D$,  $U_s$ and $D/U_s$, respectively. The dimensionless fluid velocity averaged over the circular cross-section reads
\begin{equation}
U(t) = U_s\cdot[1 + \A\cdot \textrm{sin}(2\pi \cdot t/T)],
\label{eq:Re}
\end{equation}
where $T=\pi \Rey_s/(2\Wo^2)$ is the dimensionless pulsation period. In oscillatory flow, we rendered lengths, velocities and time dimensionless with $D$, $U_o$ and $D/U_o$, respectively. {The  Navier--Stokes equations linearized about the Sexl-Womersely solution \citep{Womersley55}, and the corresponding adjoint equations,} were discretized in cylindrical coordinates $(r, \theta, z)$ with a Chebyshev collocation method for each Fourier mode $(k, m)$, where $k$ and $m$ are the axial and azimuthal wavenumbers of the perturbation, respectively. A second-order projection scheme was used to integrate the equations in time \citep[see][for details of the method]{xu2020}. {The code was validated against the Floquet analysis of \citet{Thomas12}}. {In this study, we used $N=96$ radial points and time step $\Delta t=0.002$} (convergence was checked at selected parameter values with $\Delta t=0.0005$ and  $N=128$). The optimal transient energy growth of a disturbance $\mathbf{u'}_{km}$ with wavenumbers $(k, m)$ was computed as
	\begin{equation}
	G_{km}(t_0,\tau) = \max\limits_{||\mathbf{u'}_{km}(t_0)||_2\neq0} \; \frac{E_{km}(t_0+\tau)}{E_{km}(t_0)},
	\label{eq:G}
	\end{equation}
with the adjoint method of \citet{Barkley08}. Here $E_{km}(t)$ is the kinetic energy of the disturbance at time $t$,  $t_0$ the time (phase) at which the perturbation is applied and $t_f$ the point at which the growth is evaluated ($\tau=t_f-t_0$ is the perturbation evolution time). 
We varied $\Rey_s$, $\A$, and $\Wo$ independently and for each set of parameter values the maximum transient growth $G_\text{max}$ was optimized  over $t_0$, $\tau$, $k$ and $m$. We found that in most regimes the optimal azimuthal wavenumber is $m=1$ (except for some regimes of oscillatory flow). In all the results shown below  $m=1$ unless otherwise specified. 

\section{Dynamics of the optimal disturbance}

\begin{figure}
	\centering
	\includegraphics[width=1\textwidth, trim={0cm 0cm 0cm 0cm}, clip]{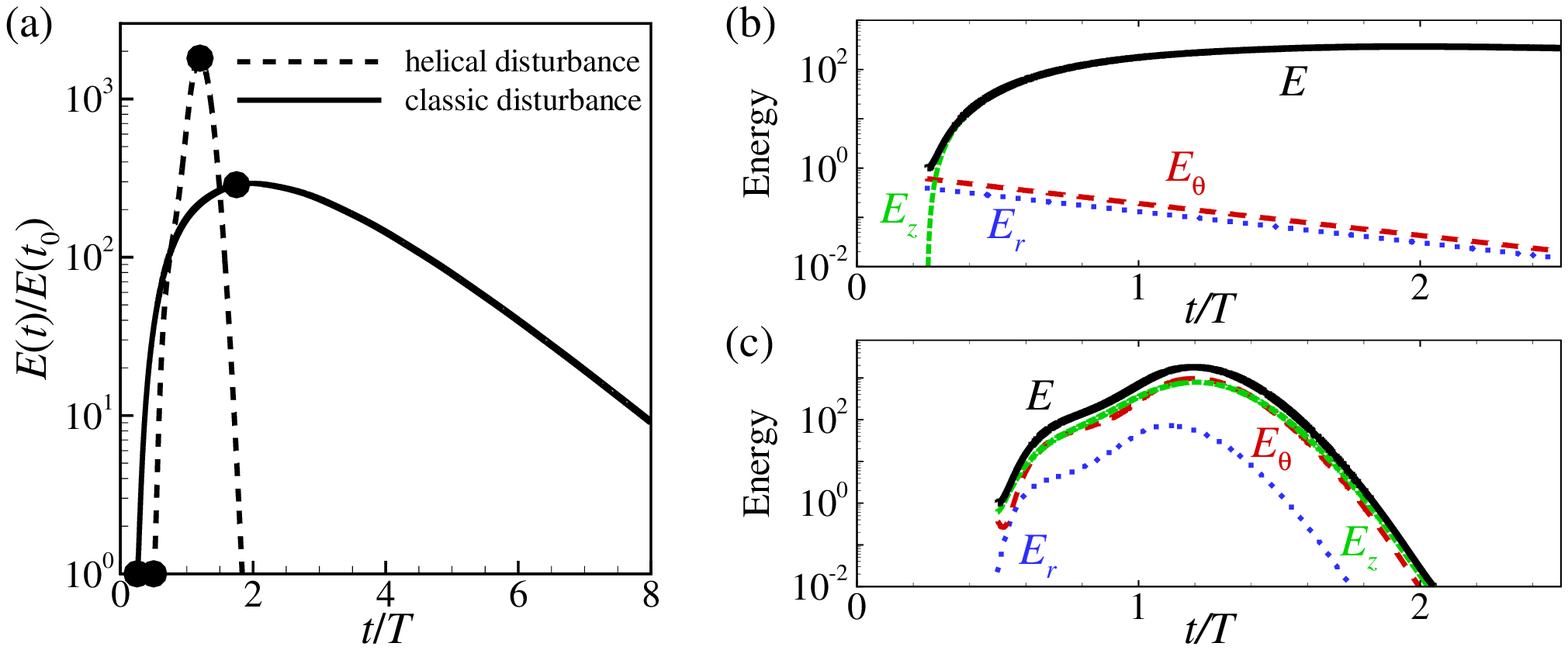}\\
	\includegraphics[width=1\textwidth, trim={0cm 0cm 0cm 0cm}, clip]{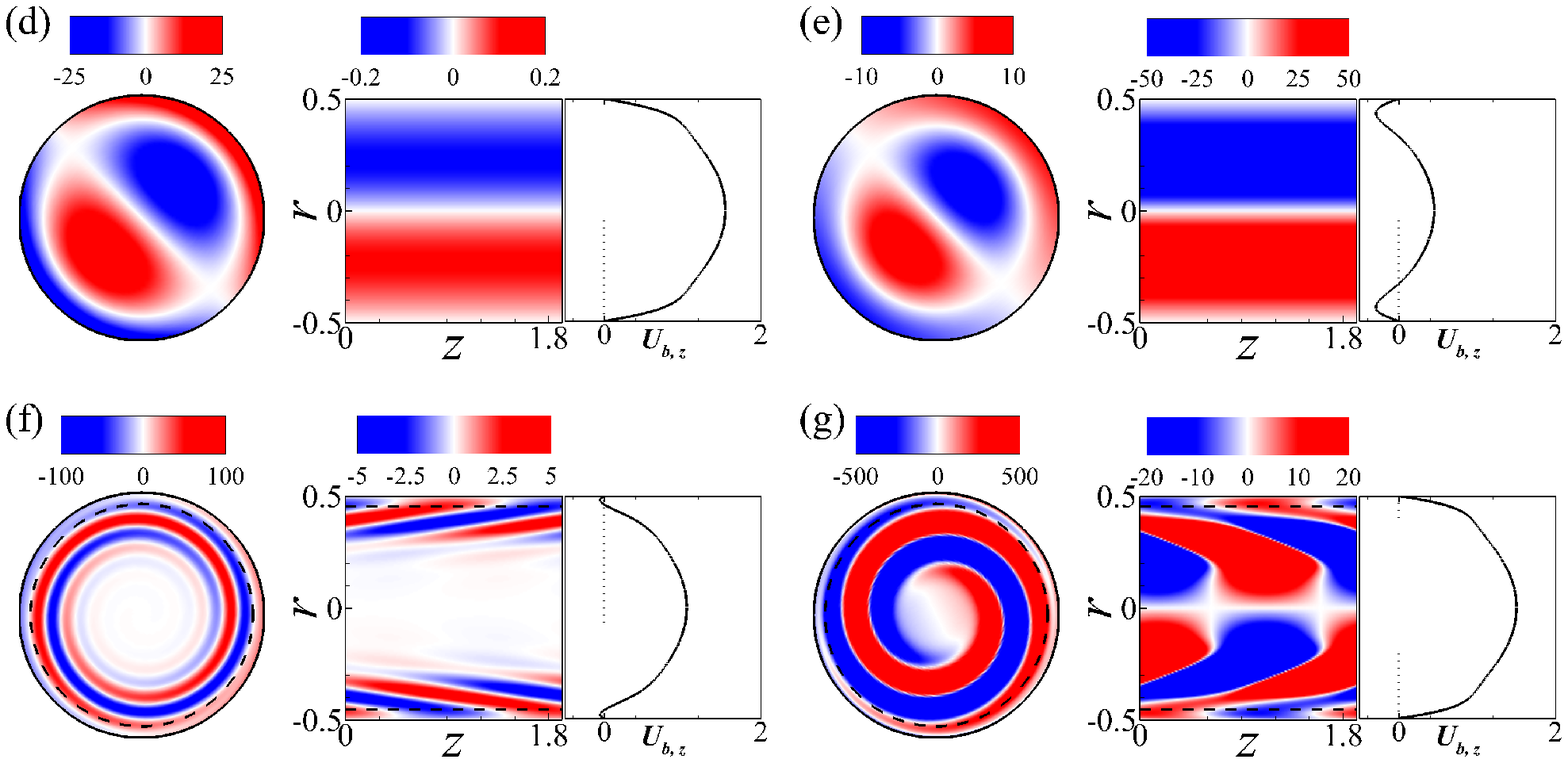}\\	
	\caption{\label{fig:Re2000Wo15A1_OptimalPerturbation} (a) Time series of the kinetic energy $E(t)/E(t_0)$ of the optimal helical, $(k,m)=(3.24,1)$, and classic, $(k,m)=(0,1)$, disturbances at $(Re_s,\A,\Wo)=(2000,1,15)$. (b)--(c) Time series of kinetic energy contribution of each velocity component  for the classic (b) and helical (c) disturbances.  (d)--(e) Contours of stream-wise vorticity (on a $r$-$\theta$ cross-section) and of stream-wise velocity (on a $z$-$r$ cross-section) of the classic disturbance, and the corresponding base flow profile ${\bm U}_{b,z}$ at $t_0/T=0.25$ (d) and $t_f/T=1.75$; see supplementary movie 1 for an animation of the disturbance dynamics  (e). (f)--(g) The same as (d)--(e), but for the helical disturbance at $t_0/T=0.5$ and $t_f/T=1.2$; see also supplementary movie 2. The dashed line denotes the Stokes-layer thickness.}
\end{figure}

The temporal evolution of the optimal disturbance's energy at $\Rey_s=2000$, $\A=1$ and $\Wo=15$  is shown as a dashed line in figure~\ref{fig:Re2000Wo15A1_OptimalPerturbation}(a). At these parameter values the optimal disturbance has a helical structure with $(k,m)=(3.24,1)$ and is localized at the outer half of the pipe (exceeding the Stokes layer thickness), see figure~\ref{fig:Re2000Wo15A1_OptimalPerturbation}(f)--(g). The optimal point to disturb is during the deceleration phase, at $t_0/T=0.5$, whereas the maximum amplification is reached during the acceleration phase, at $t_f/T=1.2$. The classic $k=0$, $m=1$  optimal disturbance of steady pipe flow is also amplified significantly in this case, albeit an order of magnitude less than the helical one. The classic disturbance initially consists of stream-wise vortices, as shown in figure~\ref{fig:Re2000Wo15A1_OptimalPerturbation}(d)--(e), and the energy is subsequently transferred to the stream-wise velocity components, while the cross-stream components decay monotonically, see figure~\ref{fig:Re2000Wo15A1_OptimalPerturbation}(b) and supplementary movie 1. Overall the classic perturbation's behavior appears to be rather insensitive to the change in flow profile throughout the cycle and the decay is very slow. 

The kinetic energy of the optimal helical perturbation is mostly distributed in the stream-wise and azimuthal components, which self-amplify rapidly during the deceleration phase and a bit slower during the subsequent acceleration phase, see figure~\ref{fig:Re2000Wo15A1_OptimalPerturbation}(c). Initially the disturbance spirals clock-wisely towards the pipe center while leaning against the background shear profile, see figure~\ref{fig:Re2000Wo15A1_OptimalPerturbation}(f) and supplementary movie 2. As the energy grows, the perturbation switches the spiraling direction and is tilted by the shear until it aligns with it and the disturbance finally decays, see figure~\ref{fig:Re2000Wo15A1_OptimalPerturbation}(g). This is reminiscent of the Orr mechanism \citep{Orr1907}. However, approximately $96\%$ of the kinetic energy is shared in equal parts between the azimuthal and stream-wise components, which indicates a strong three-dimensional effect, distinct from the two-dimensional Orr mechanism reported for many flows \citep[see e.g.][]{boyd1983,Farrell88,Maretzke2014, Biau16}. 

{\citet{vonkerczek1982,Tsigklifis17} studied the stability of pulsatile channel flows and reported large \emph{modal} growth of disturbances at low $\Wo$, which also sets in during the deceleration phase and reaches the maximum at the end of the period of pulsation. These authors associated the modal energy growth with the inflectional velocity profiles occurring in the deceleration phase. Here, we found that at the optimal perturbation point $t_0$, the velocity profile has also inflection points, see figure~\ref{fig:Re2000Wo15A1_OptimalPerturbation}(f).  We froze this flow profile and did a modal (eigenvalue) analysis to obtain the most dangerous mode. This mode was subsequently used as initial condition for the linearized Navier--Stokes equations and exhibited large energy growth (albeit an order of magnitude lower than the non-modal optimal). We found similar results in additional runs at $\Wo=5$ and $8$, suggesting that inflection points and modal mechanisms may also play an important role in pulsatile pipe flow.}

\begin{figure}
	\centering
	\includegraphics[width=0.32\textwidth, trim={0cm 0cm 3cm 0cm}, clip]{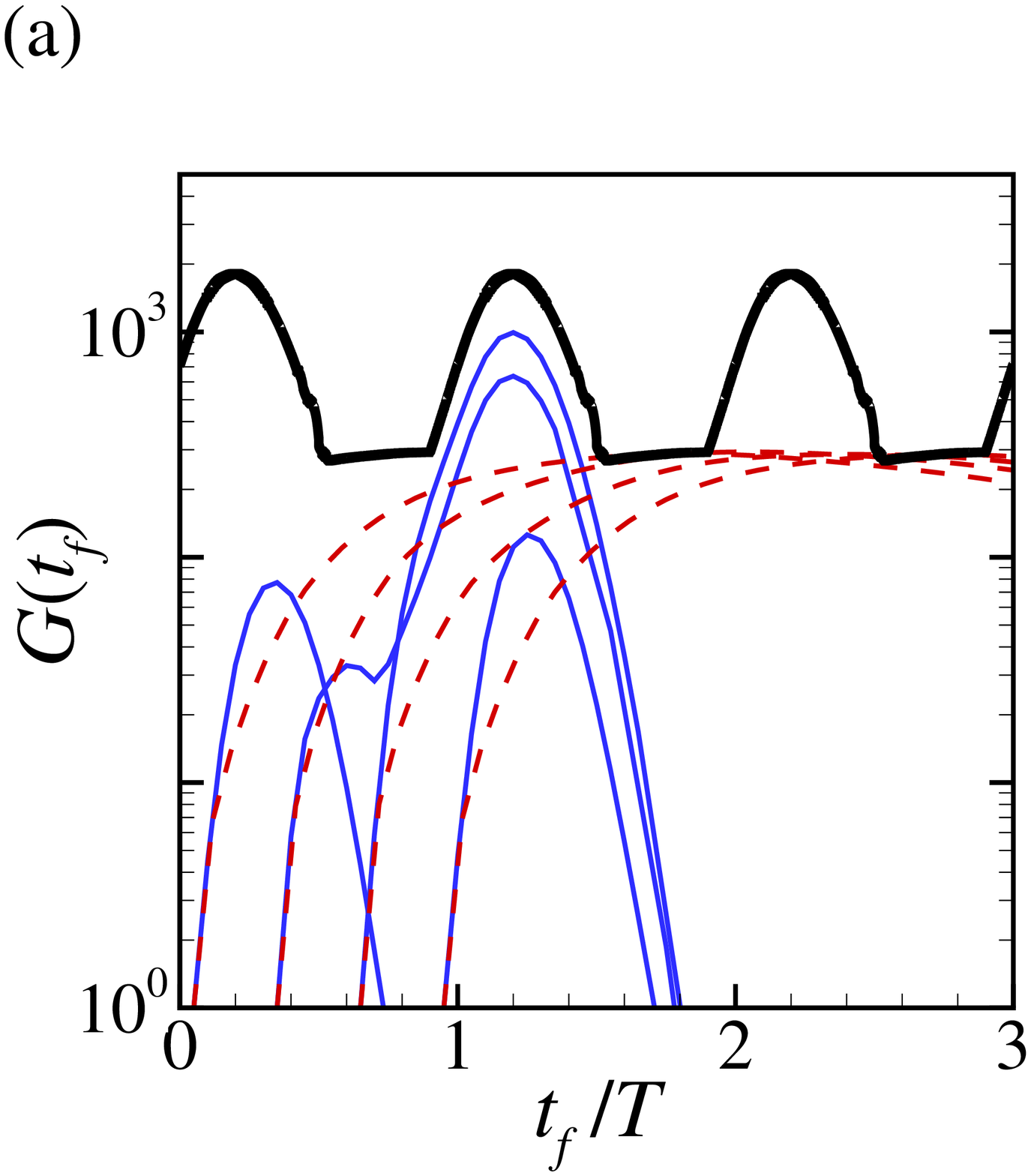}
	\includegraphics[width=0.32\textwidth, trim={0cm 0cm 3cm 0cm}, clip]{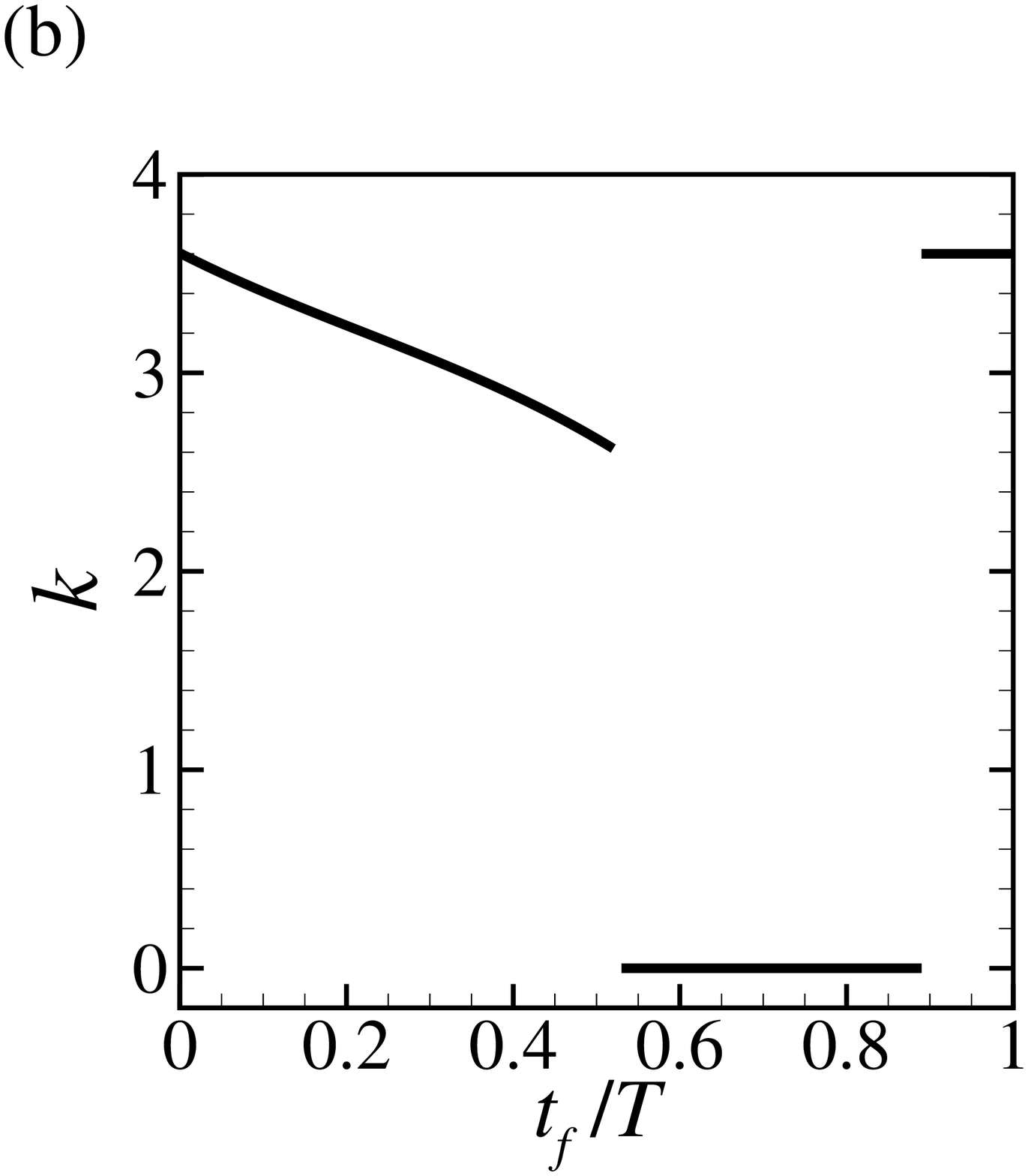}
	\includegraphics[width=0.32\textwidth, trim={0cm 0cm 3cm 0cm}, clip]{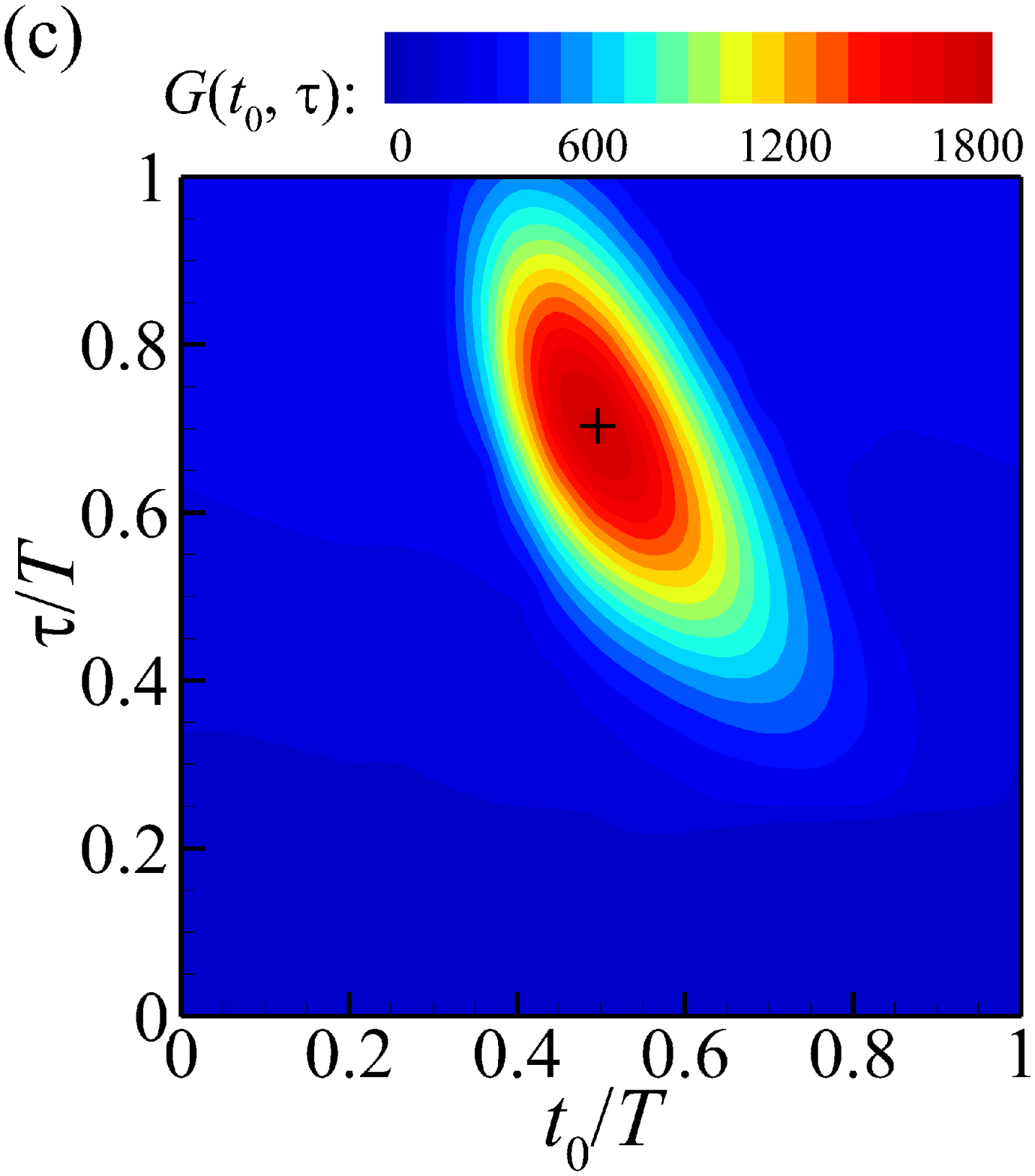}
	\caption{\label{fig:Re2000Wo15A1} Transient growth at $(Re_s,\A,\Wo)=(2000,1,15)$: (a) The red dashed lines denote the temporal evolution of the optimal classic disturbances $(k,m)=(0,1)$ for four different initial times $t_0$, whereas the blue solid lines correspond to the optimal helical disturbances $(k,m)=(3.24,1)$ initialized at the same $t_0$. The thick black line is the maximum gain $G(t_f)$ that can be achieved at a given time $t_f$ (optimized over $k$, $m$ and $t_0$ disturbances). (b) Dependence of the optimal axial wavenumber $k$ (associated to the thick line of a) on $t_f$. (c) Colormap of the maximum gain $G(t_0,\tau)$ (optimized over $k$ and $m$) in the $t_0-\tau$ plane. The black cross marks the maximum of $G$.}
\end{figure}

The black thick line in figure~\ref{fig:Re2000Wo15A1}(a) depicts the maximum energy amplification $G(t_f)$ over all $(k,m)$ and initial disturbances time $t=t_0$. The maximum amplification is reached during the acceleration phase via helical disturbances, as shown in figure~\ref{fig:Re2000Wo15A1}(b), whereas the classic disturbance achieves larger growth only during the second half of the deceleration phase. The colormap of figure~\ref{fig:Re2000Wo15A1}(c) shows that the optimal time to perturb the flow is during middle of the deceleration phase ($t_0/T\approx0.5$); perturbing during the acceleration phase leads to much lower growth (yielded by the classic disturbance during the deceleration phase). Clearly the helical mechanism is efficient only in the deceleration phase. The time needed to reach the maximum growth is about $\tau\approx 3/4T$, which explains why the maximum growth occurs during the acceleration phase.   

\section{Parametric study of transient growth}

\begin{figure}
	\centering
	\includegraphics[width=0.32\textwidth, trim={0cm 0cm 1.5cm 0cm}, clip]{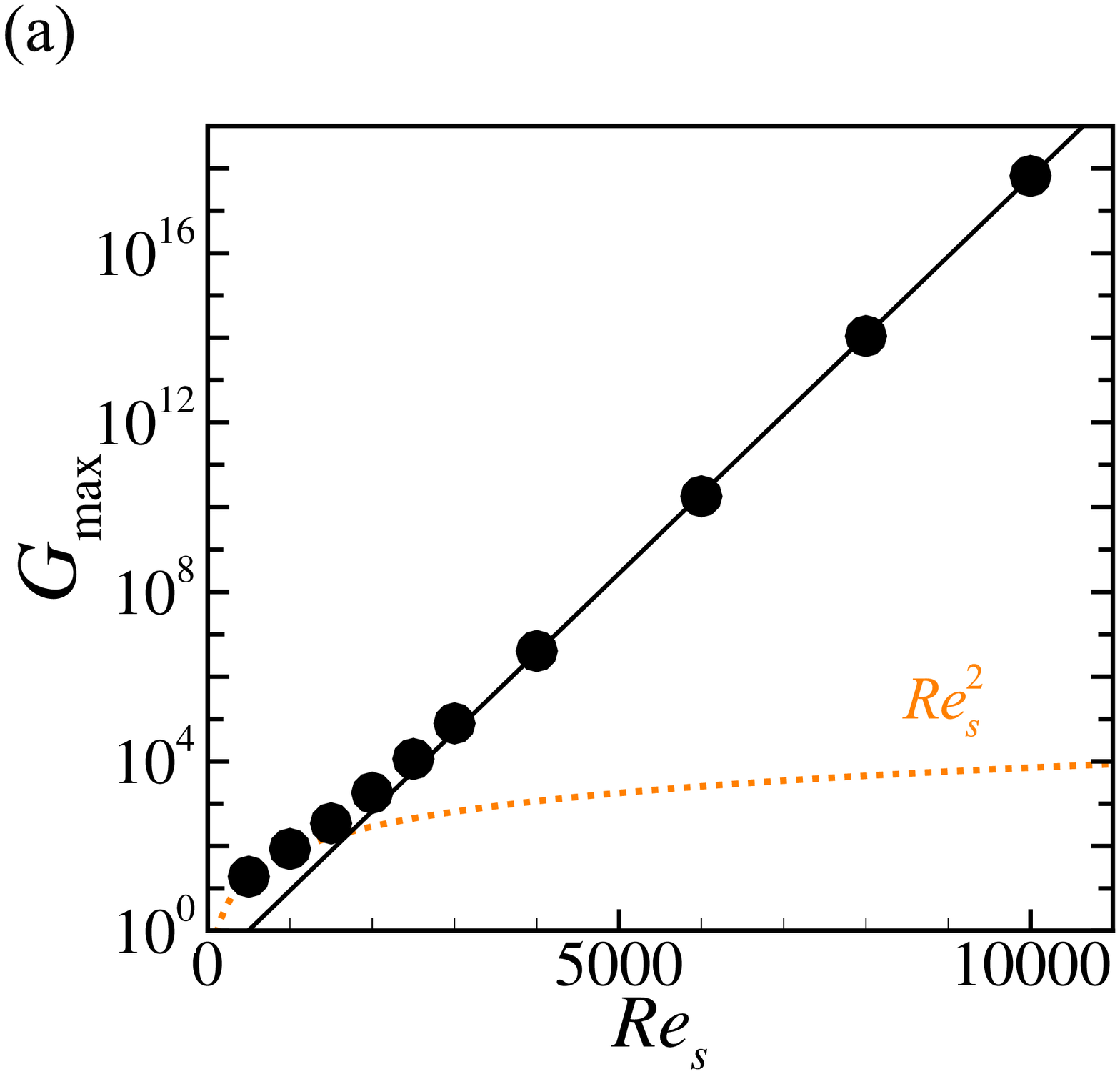}
	\includegraphics[width=0.32\textwidth, trim={-0.5cm 0cm 2cm 0cm}, clip]{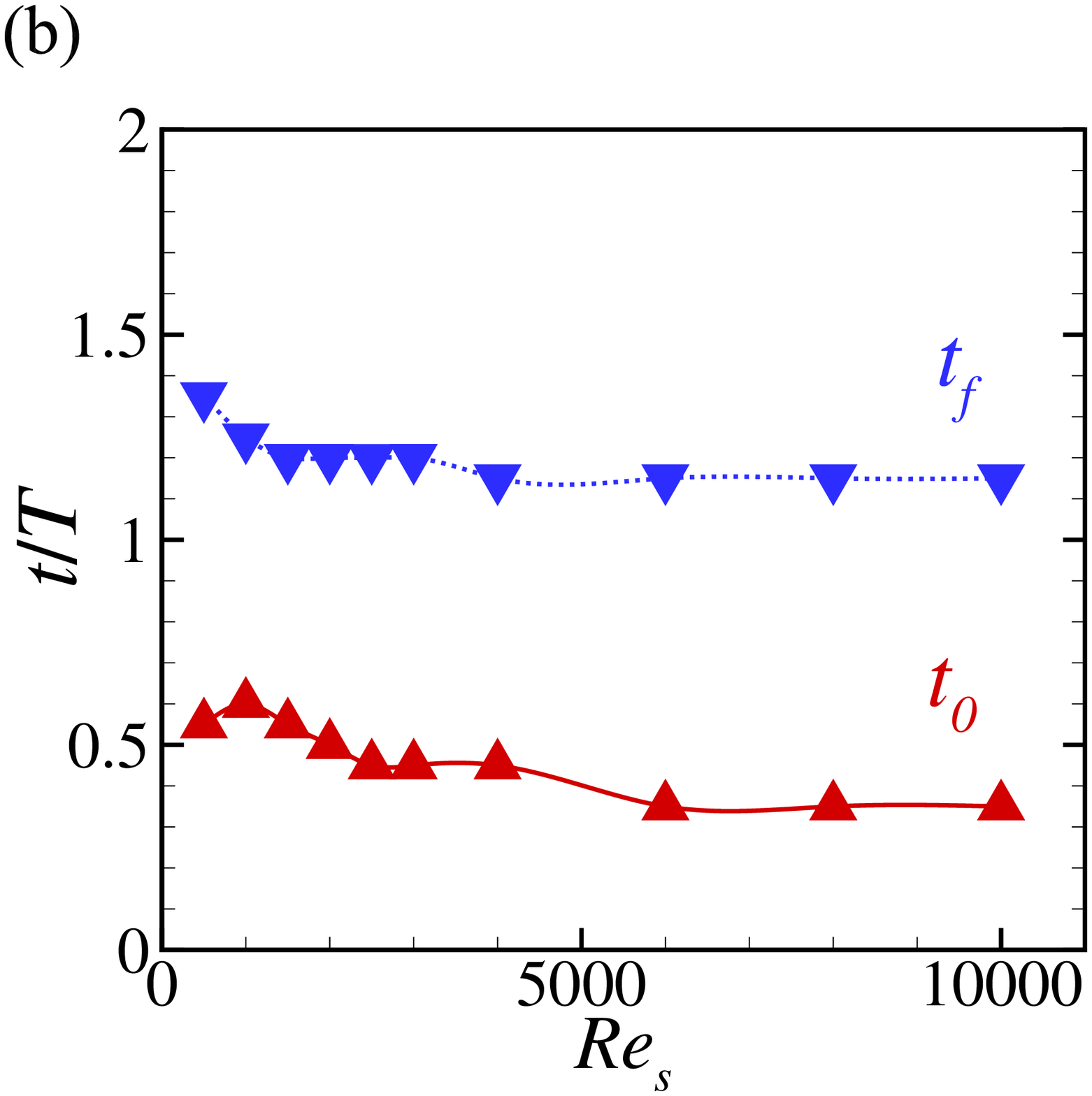}
	\includegraphics[width=0.32\textwidth, trim={0cm 0cm 1.5cm 0cm}, clip]{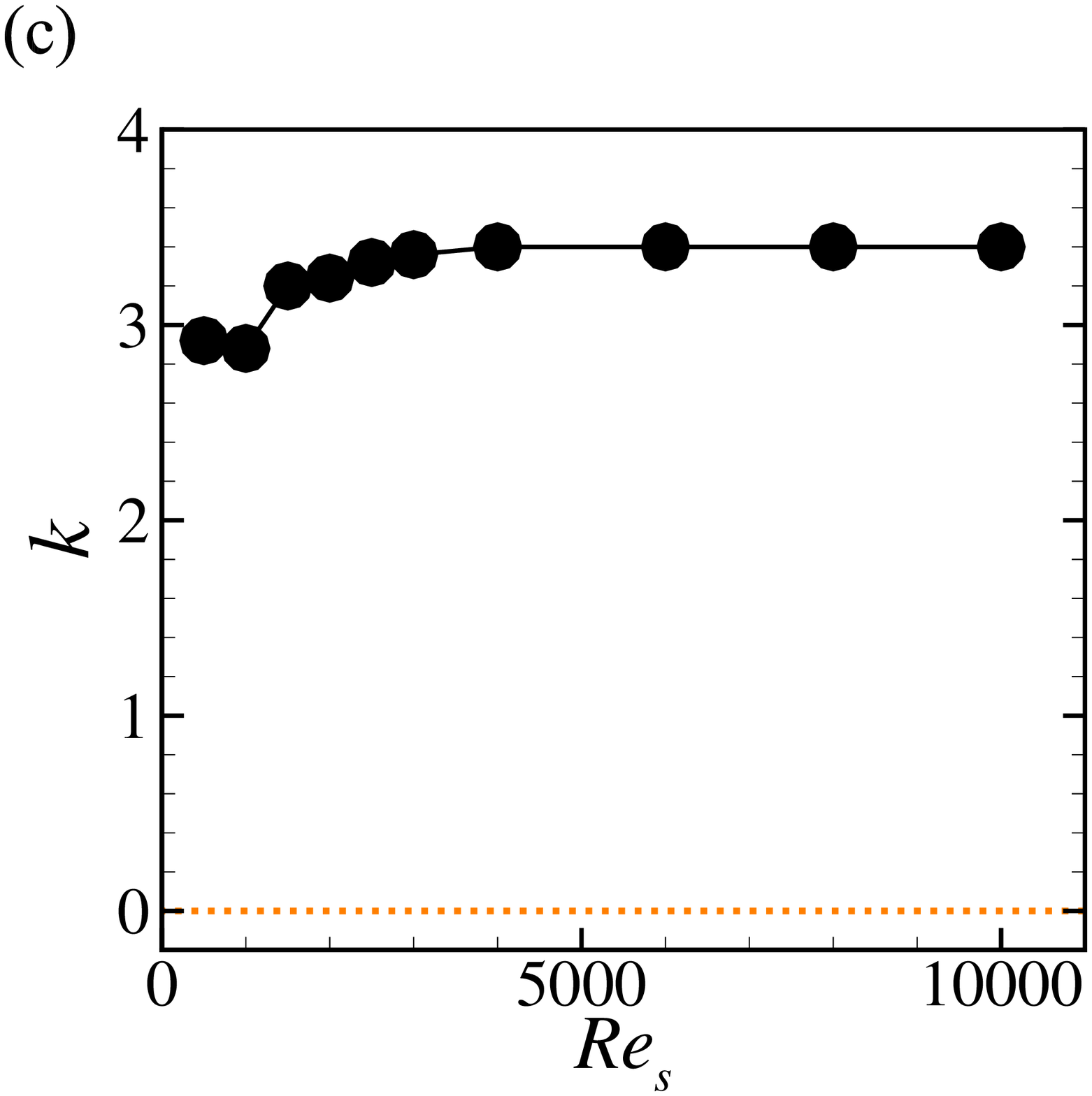}
	\caption{\label{fig:Re_dependence} Optimal transient growth as a function of $\Rey_s$ for $(\A,\Wo)=(1,15)$. (a) The symbols show the computed maximum energy gain. The black solid line $G_\text{max}=\exp(\Rey_s/232-2.12)$ is a fit to the data for $\Rey_s\geq4000$. The orange dotted line $G_\text{max}=2.72\,\Rey_s^2$ is a fit to data for $(\A,\Wo)=(1,25)$. (b)--(c) Optimal initial and final disturbance times and axial wavenumber.}
\end{figure}

We show in figure~\ref{fig:Re_dependence}(a) that keeping $(\A,\Wo)=(1,15)$ fixed and increasing the Reynolds number leads to an approximately exponential increase of the energy gain $G_\text{max}$. Independently of the Reynolds number, the maximum energy amplification occurs always during flow acceleration with helical disturbances introduced during the deceleration phase, see figure~\ref{fig:Re_dependence}(b). As shown in figure~\ref{fig:Re_dependence}(c) an asymptotic behavior is approached for $\Rey_s\gtrsim 5000$, with $k\approx3.4$, $t_f/T\approx1.15$ and $t_0/T\approx0.35$. 

\begin{figure}
	\centering
	\includegraphics[width=0.32\textwidth, trim={0cm 0cm 1.5cm 0cm}, clip]{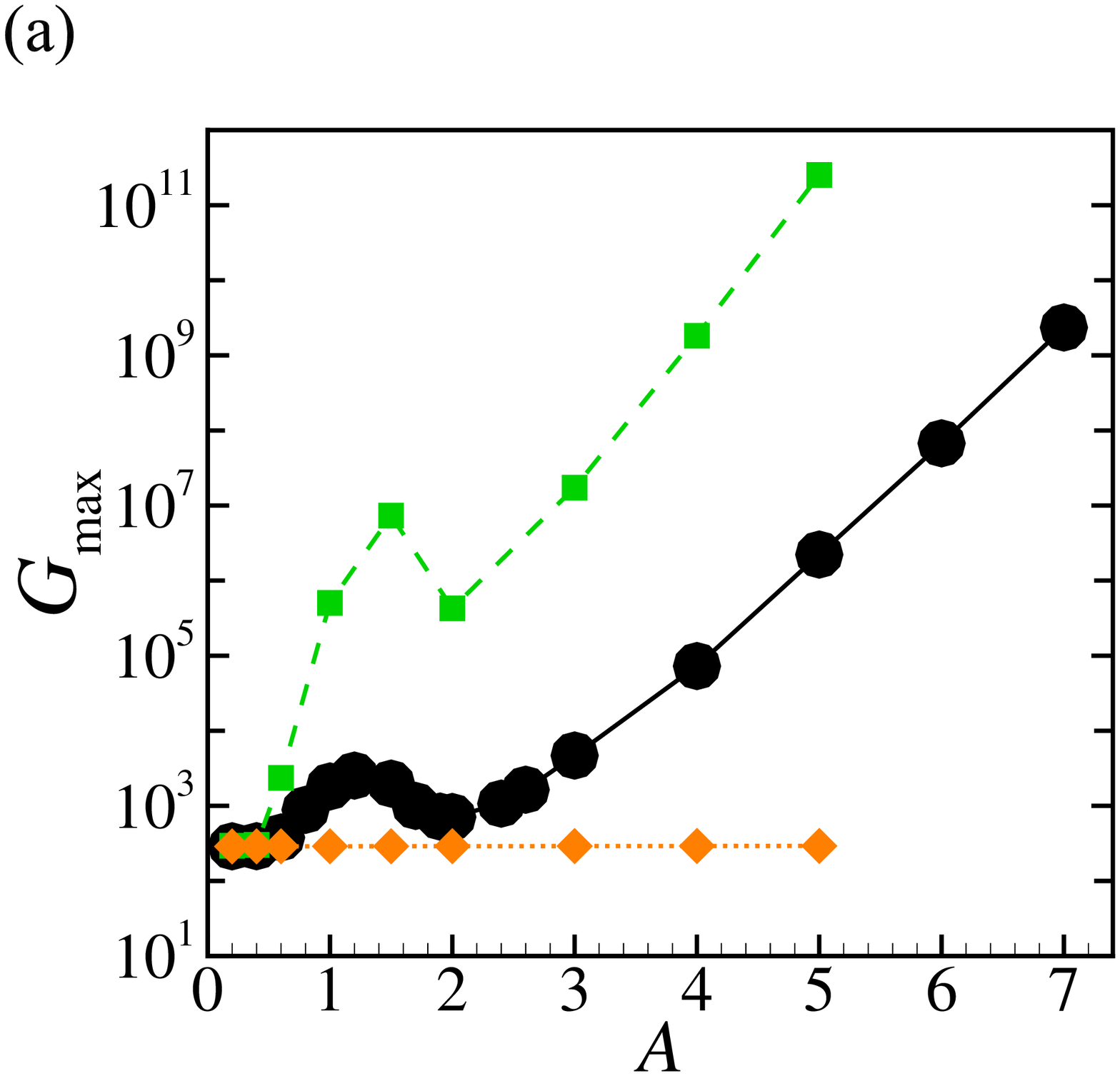}
	\includegraphics[width=0.32\textwidth, trim={-0.5cm 0cm 2cm 0cm}, clip]{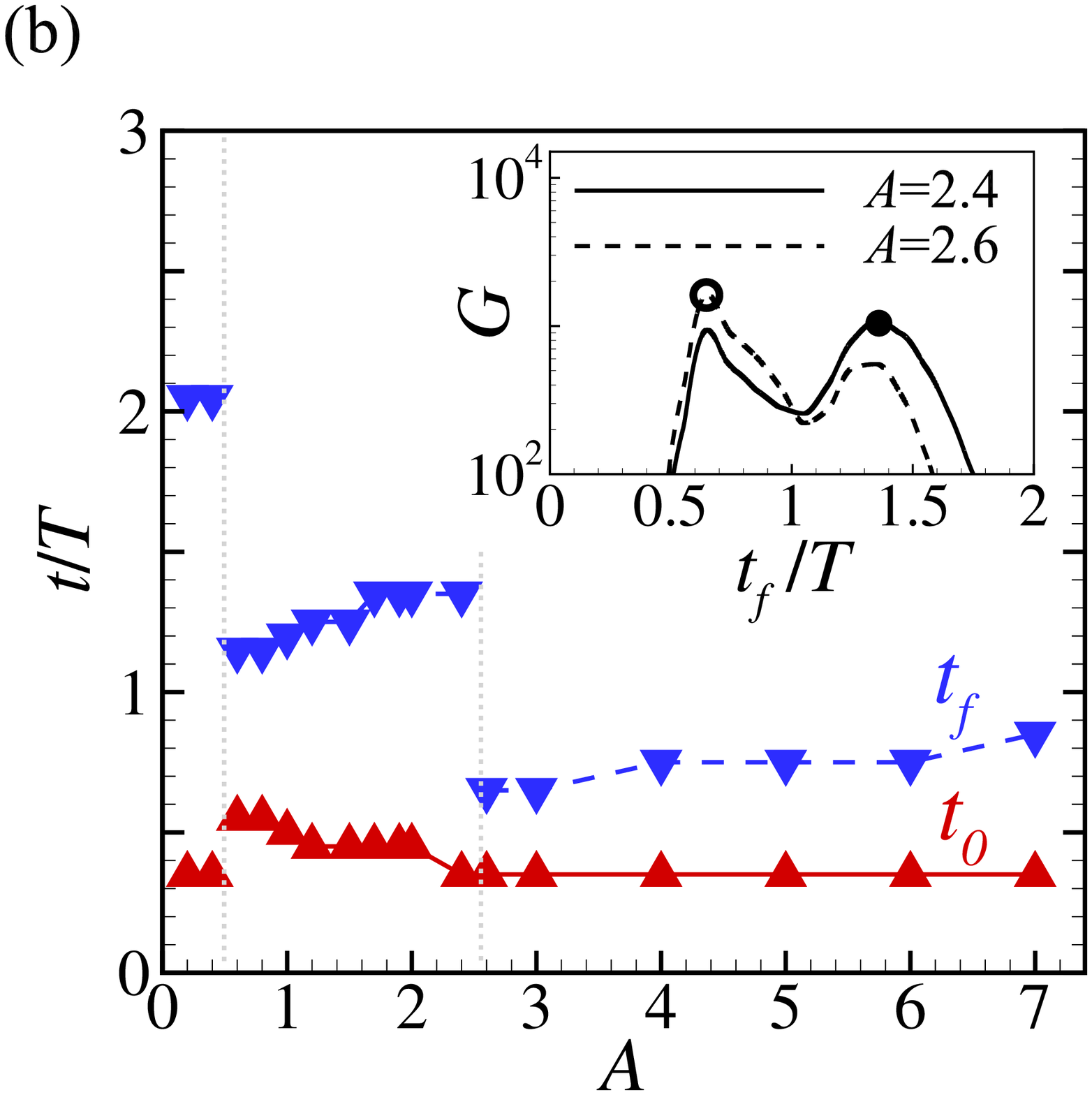}
	\includegraphics[width=0.32\textwidth, trim={0cm 0cm 1.5cm 0cm}, clip]{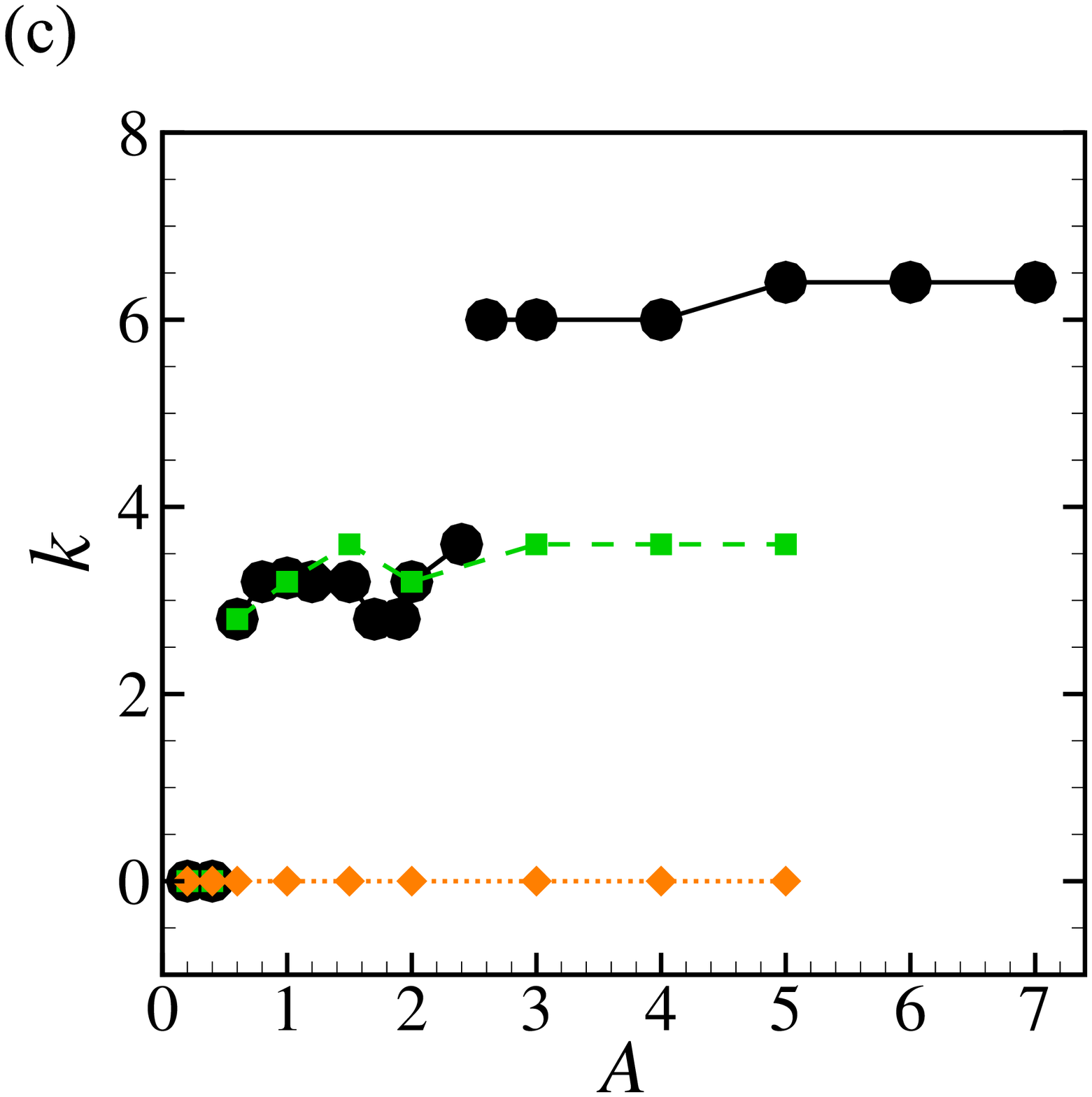}
	\caption{\label{fig:amplitude_dependence} Optimal transient growth as a function of $\A$ for $(\Rey_s,\Wo)=(2000,15)$. (a) The black circles show the computed maximum energy gain, the green squares and the orange diamonds are for $(\Rey_s,\Wo)=(2000,10)$ and $(\Rey_s,\Wo)=(2000,20)$, respectively. (b)--(c) Optimal initial and final disturbance times and axial wavenumber. The maximum gain as a function of $t_f/T$ is shown in the inset in (b)  for $\A=2.4$ and $2.6$ and illustrates the competition between two distinct different disturbances. }
\end{figure}

The effect of the pulsation amplitude $\A$, whilst keeping  $(\Rey_s,\Wo)=(2000,15)$ fixed is illustrated in figure~\ref{fig:amplitude_dependence}.  For steady pipe flow ($\A=0$), \citet{Meseguer03} obtained $G_\text{max}\approx288.7$, $\tau\approx 24.5$ and $(k,m)=(0,1)$, which is in line with our results were for pulsatile flow of low amplitude $A\lesssim0.55$. Helical disturbances dominate thereafter {(see the black line in figure~\ref{fig:amplitude_dependence}a)}, but their energy gain does not grow monotonically (exponentially) with the amplitude until $\A\gtrsim4$. For intermediate amplitudes there is competition between two distinct types of helical disturbances, as shown in figures~\ref{fig:amplitude_dependence}(b). For $0.55\lesssim\A\lesssim2.5$, the dominant helical disturbance is similar to that examined in detail in figures~\ref{fig:Re2000Wo15A1}--\ref{fig:Re_dependence}, whereas for $\A\gtrsim 2.5$ another helical disturbance, with much shorter axial length and shorter evolution time $\tau=t_f-t_0$, takes over. Both types of helical disturbances are triggered in the deceleration phase. Similar results were obtained for lower frequency, at $(\Rey_s,\Wo)=(2000,10)$, see the green lines in figures~\ref{fig:amplitude_dependence}(a) and (c). Here the maximum growth $G$ is over two orders of magnitude larger and helical disturbances dominate earlier (already for $A\gtrsim 0.41$). From the data in figures~\ref{fig:Re_dependence}(a) and \ref{fig:amplitude_dependence}(a) we conclude that the maximum energy gain of helical disturbances scales exponentially with the Reynolds number of the oscillatory flow component $\Rey_o = U_o D /\nu=A \Rey_s$.

\begin{figure}
	\centering
	\includegraphics[width=0.32\textwidth, trim={0cm 0cm 1.5cm 0cm}, clip]{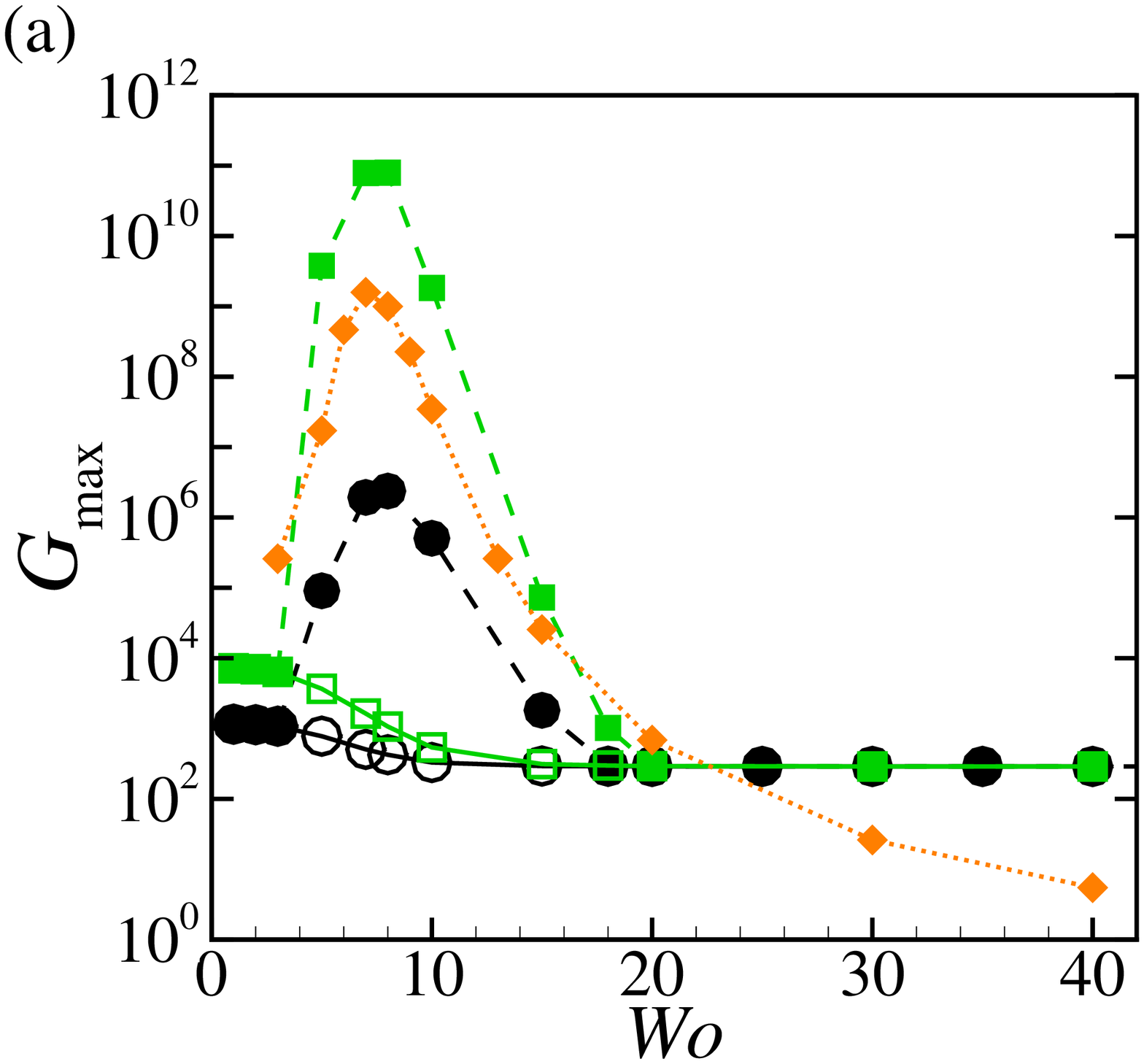}
	\includegraphics[width=0.32\textwidth, trim={0cm 0cm 1.5cm 0cm}, clip]{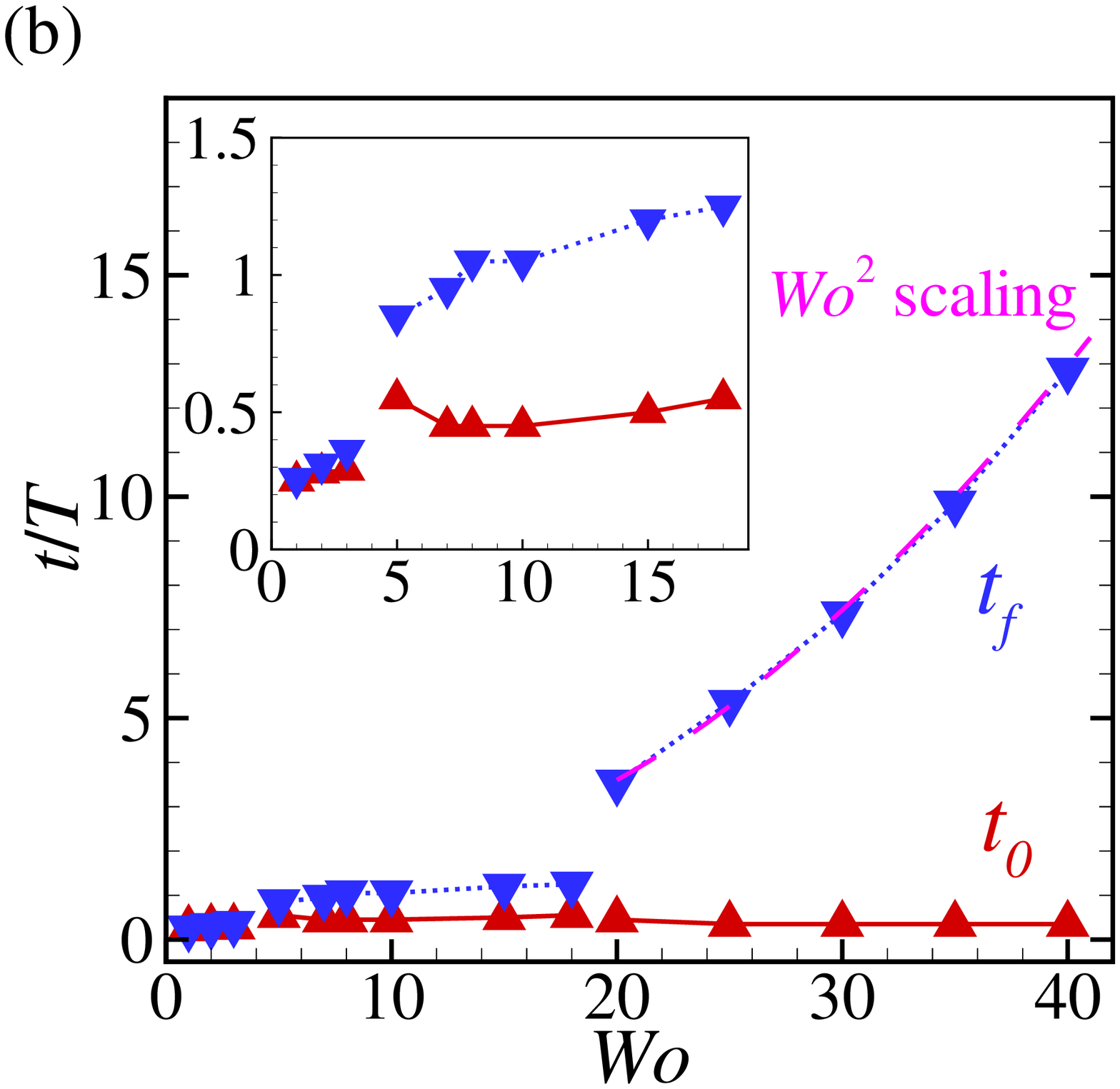}
	\includegraphics[width=0.32\textwidth, trim={0cm 0cm 1.5cm 0cm}, clip]{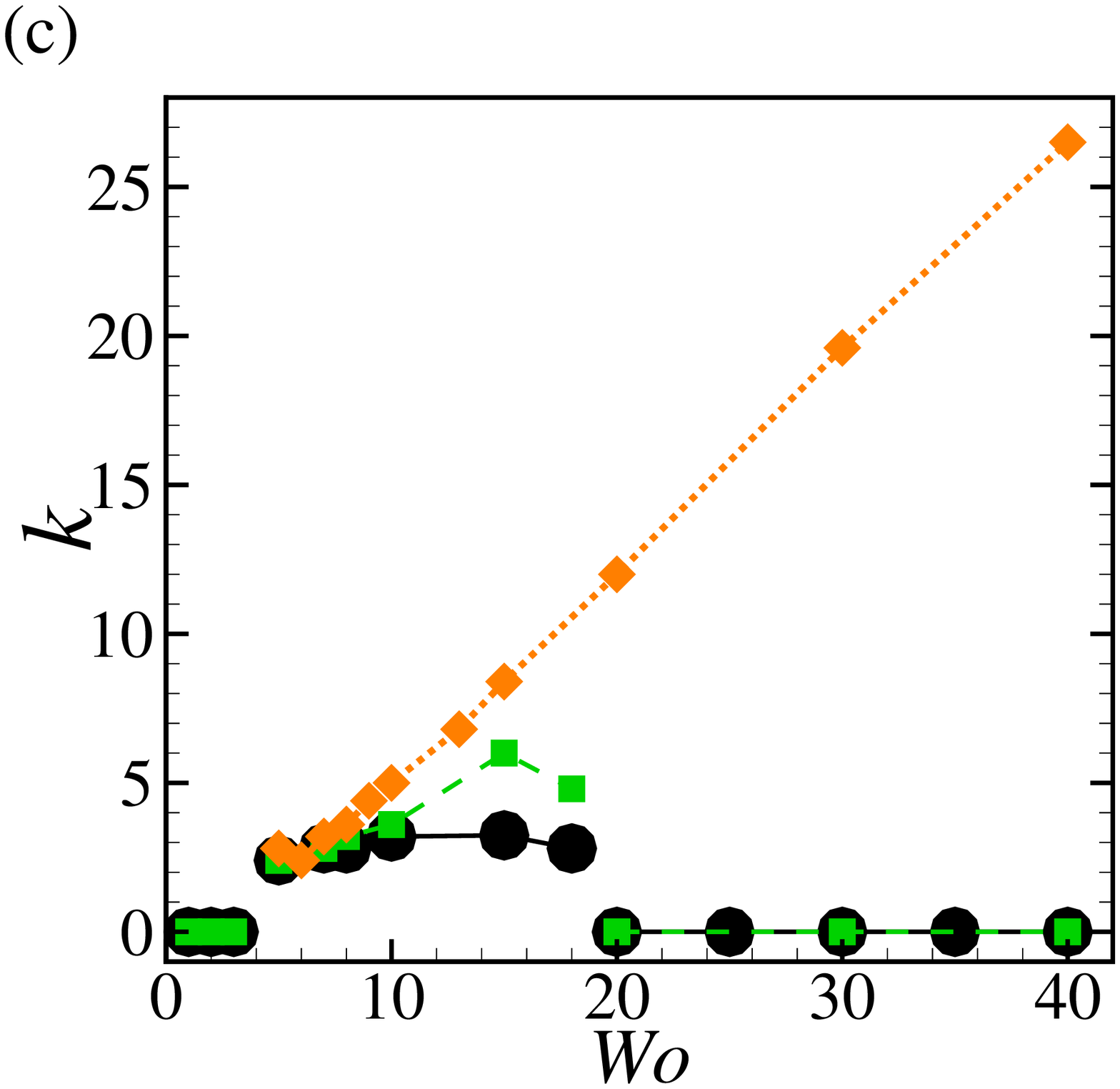}
	\caption{\label{fig:frequency_dependence} Effect of the {Womersley} number $\Wo$ on  the transient growth. (a) Maximum energy gain for $(\Rey_s,\A)=(2000,1)$ (black circles), $(\Rey_s,\A)=(2000,4)$ (green squares) and oscillatory flow at $\Rey_o=8000$ $(\Rey_s,\A)=(0,\infty)$ (orange diamonds). The hollow symbols show the optimal growth of classic disturbances $(k,m)=(0,1)$ in the regime where the helical disturbance domiates. (b) Optimal initial and final disturbance times at $(\Rey_s,\A)=(2000,1)$. The pink dashed line shows $t_f/T = (\tau_\text{steady} + t_0)/T$, where $\tau_\text{steady} $ is the optimal evolution time of the classic disturbance in steady pipe flow at $\Rey_s=2000$. {The inset shows a zoom for $\Wo\leqslant18$.} (c) Optimal axial wavenumber.}
\end{figure}

In what follows we  examine the influence of the pulsation frequency on the dominant mechanism of transient growth in more detail. We begin by focusing on $\Rey_s=2000$ and $\A=1$. The circles in figure~\ref{fig:frequency_dependence}(a) show that  for sufficiently large $\Wo\gtrsim 20$ the classic disturbance dominates and the optimal gain $G$ of steady pipe flow is recovered (exactly as for the data shown as a orange lines in figures~\ref{fig:Re_dependence} and \ref{fig:amplitude_dependence}). As shown in figure~\ref{fig:frequency_dependence}(b), the perturbation growth time $\tau\approx25$ is also indistinguishable from that of steady pipe flow.  It can be concluded that in the limit $\Wo\rightarrow \infty$ the disturbance response is solely governed by $\Rey_s$. The dynamics of the steady pipe flow is also recovered in the quasi-steady limit $\Wo\rightarrow 0$, where the classic perturbation dominates  as well. In this limit, the maximum transient growth is governed by the maximum Reynolds number $\Rey_\text{max} = (1+\A)\Rey_s=\Rey_s+\Rey_o$ and occurs for $t_0,t_f\rightarrow T/4$, see the inset in figure~\ref{fig:frequency_dependence}(b).  For instance, for $\Rey_s=2000$, $\A=1$ and $\Wo=3$ the maximum gain is $G\approx1100$, which is slightly less than the gain for steady pipe flow $G\approx1155$ at $\Rey_s=4000$. As shown in figure~\ref{fig:frequency_dependence}(c), helical disturbances dominate for intermediate $4\lesssim\Wo\lesssim18$ and exhibit a sharp maximum in gain at $\Wo\approx 7$, figure~\ref{fig:frequency_dependence}(a). The results for $\A=4$ (shown as green squares) are qualitatively similar to those for $\A=1$, but the energy gain is much larger.  

\section{Transient growth in oscillatory pipe flow} \label{sec:oscillation}

\begin{figure}
	\centering
	\includegraphics[width=0.32\textwidth, trim={0.5cm 0cm 0cm 0cm}, clip]{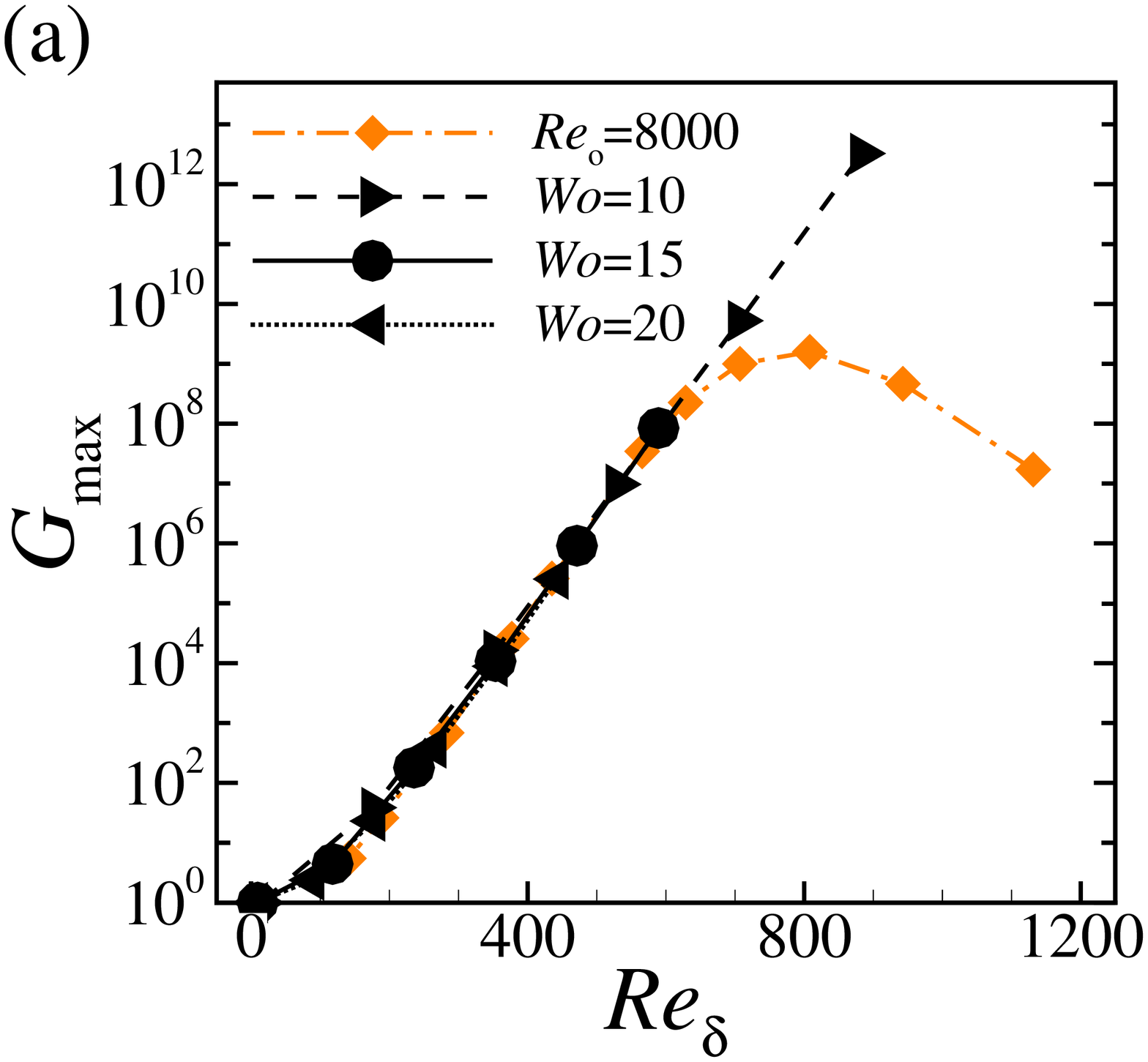}
	\includegraphics[width=0.32\textwidth, trim={0.5cm 0cm 0cm 0cm}, clip]{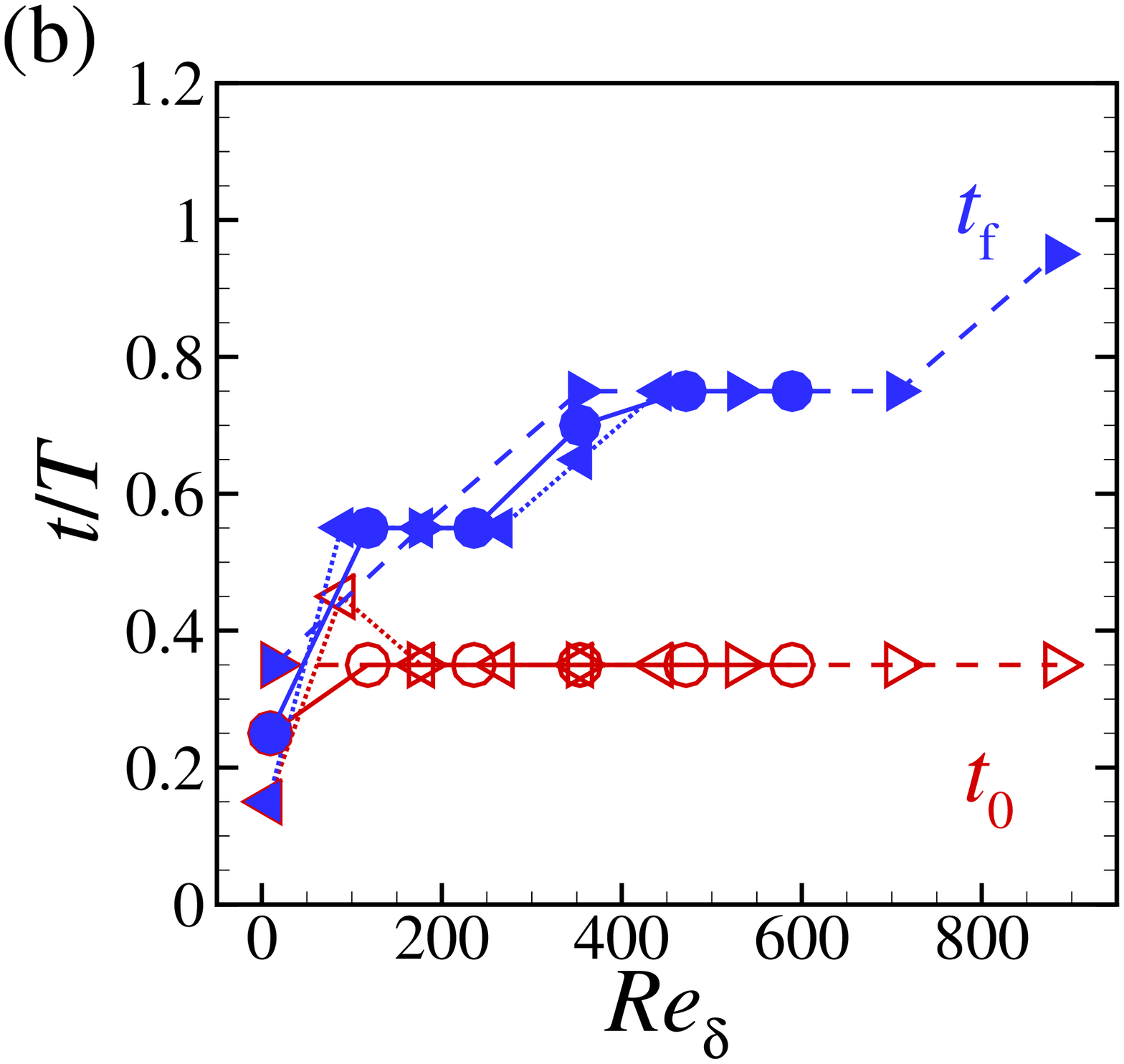}
	\includegraphics[width=0.32\textwidth, trim={0.5cm 0cm 0cm 0cm}, clip]{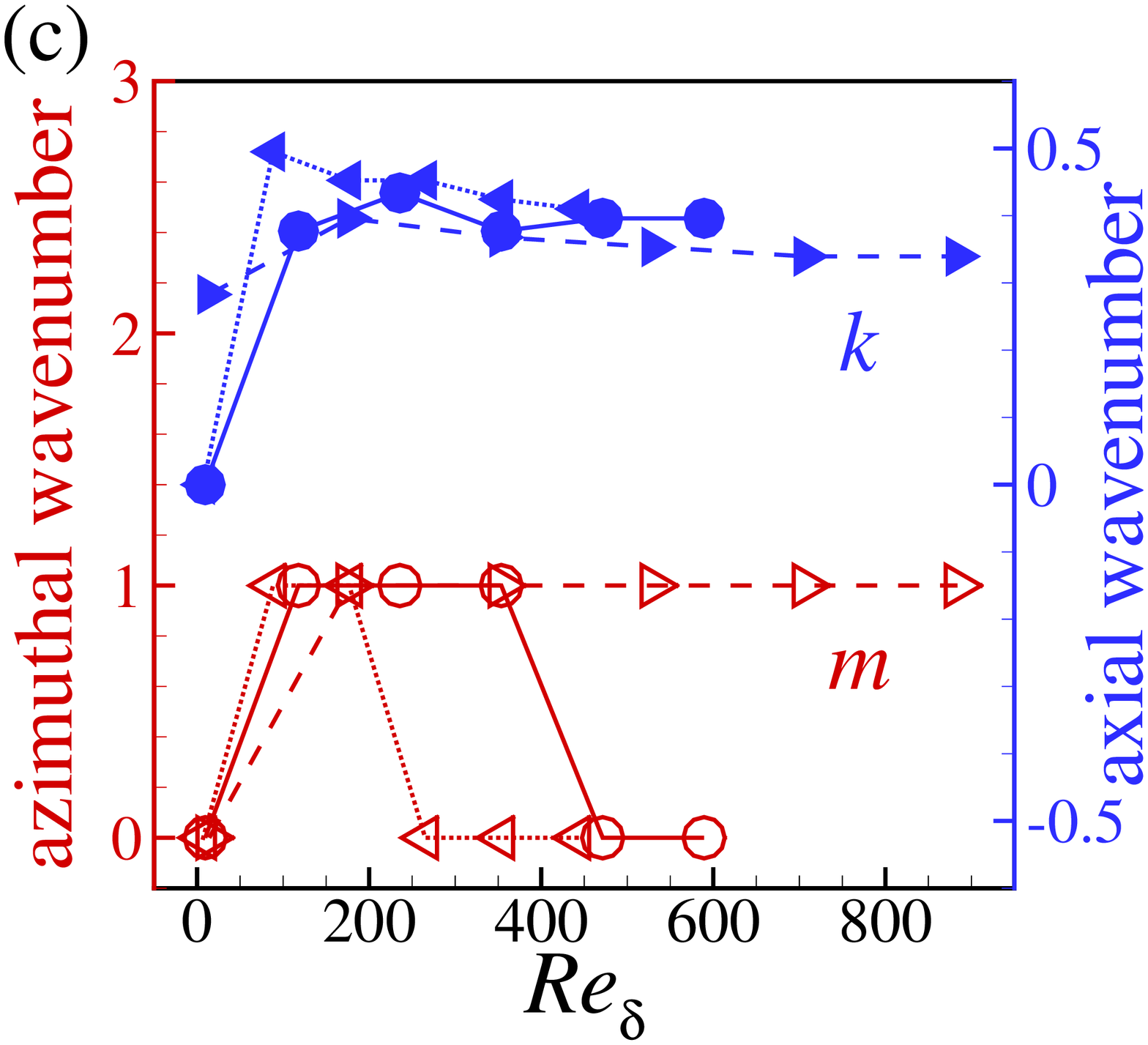}
	\includegraphics[width=1\textwidth, trim={-0.2cm 0cm 0.2cm 0cm}, clip]{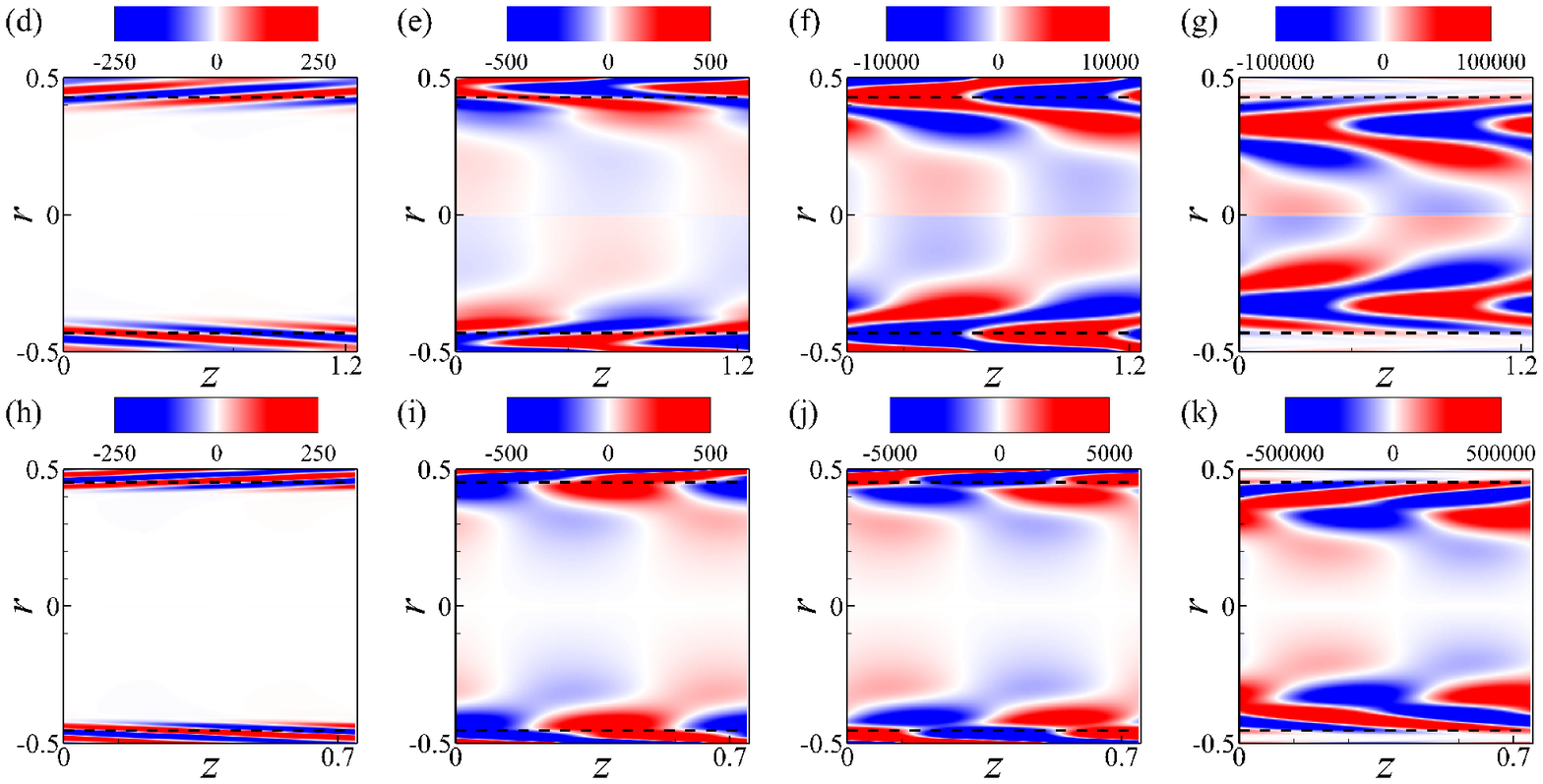}
	\caption{\label{fig:OPF} {Transient growth of disturbances in oscillatory pipe flow for $\Wo=10$, $15$ and $20$ as a function of $\Rey_{\delta}$: (a) energy amplification, (b)--(c) optimal initial and final disturbance times and optimal wavenumbers, where the axial wavenumber is scaled with $\delta^{-1}$. (d)--(g) Contours of span-wise vorticity (on a $z-r$ cross-section) of the optimal helical disturbance ($m=1$) for $\Wo=10$ and $\Rey_\delta=530$ at $t_0/T=0.35$ (d), $t/T= 0.4$ (e), $t/T=0.5$ (f) and $t_f/T=0.75$ (g); see supplementary movie 3 for an animation. (h)--(k) The same as (d)--(g), but for the optimal axisymmetric disturbance ($m=0$) for $\Wo=15$ and $\Rey_\delta=589$ at $t_0/T=0.35$ (h), $t/T= 0.4$ (i), $t/T=0.5$ (j) and $t_f/T=0.75$ (k); see supplementary movie 4.}
	}
\end{figure}

The dependence of the energy gain $G$ on the frequency for the specific case of oscillatory flow  at $\Rey_o=8000$ is shown as orange diamonds in figure~\ref{fig:frequency_dependence}(a) and follows the trend of pulsatile flow. In figure~\ref{fig:OPF}(a) these data are shown as a function of the Reynolds number of the Stokes layer $\Rey_{\delta}=U_o\delta/\nu=\Rey_o/(\sqrt{2}\Wo)$, together with three additional sets for $\Wo=10$, $15$ and $20$ covering wide ranges of $\Rey_o$. At  low $\Rey_{\delta}$, the maximum gain $G=1$ is reached for $t_0=t_f$, implying that all perturbations decay monotonously, and the disturbance with ($k=0, m=0$) is the least damped, see figure~\ref{fig:OPF}(b)--(c). This can also be seen in figure~\ref{fig:frequency_dependence}(a), where the  limit $G\rightarrow 1$ is reached by the orange data points as {$\Wo\rightarrow\infty$} (and hence $\Rey_{\delta}\rightarrow0$ for constant $\Rey_o=8000$). For $\Rey_{\delta} \gtrsim 100$ the energy gain increases exponentially with $\Rey_{\delta}$ and depends only weakly on the {Womersley} number. All data sets shown in figure~\ref{fig:OPF}(a) exhibit excellent collapse, except for the data set at $\Rey_o=8000$, which deviates from the exponential scaling for $\Rey_\delta \gtrsim 630$ (corresponding to $\Wo\lesssim9$, i.e.\ near and below the peak in figure~\ref{fig:frequency_dependence}a). The optimal point to disturb the flow is during the deceleration phase at $t_0/T\approx0.35$ (as in figure~\ref{fig:Re_dependence}b at sufficiently high $\Rey_o$). The maximum energy gain is attained toward the end of the deceleration phase at low $\Rey_{\delta}$, and moves progressively into the acceleration phase as $\Rey_{\delta}$ increases. 

Snapshots of the span-wise vorticity of the optimal perturbation are shown in figure~\ref{fig:OPF}(d)--(g) for $\Wo=10$ and $\Rey_\delta=530$. Initially, the perturbation leans against the background shear and then tilts to align with the stream-wise direction as it propagates radially inwards (see supplementary movie 3). The helical structure and dynamics of the disturbance are very similar to that shown in figure~\ref{fig:Re2000Wo15A1_OptimalPerturbation}(f), but here the disturbance is initially more confined to the Stokes layer. In figure~\ref{fig:OPF}(h)--(k) and supplementary movie 4 we show that at higher $\Wo=15$, the optimal disturbance is very similar in dynamics, but is axisymmetric (see also figure~\ref{fig:OPF}(c)). The energy growth occurring here is as in the classic Orr mechanism, with the only difference that the perturbation travels radially inwards as it changes the tilt direction. We stress that despite this difference in the disturbance geometry (helical versus axisymmetric), the energy gain, the point of disturbance and the point of maximum gain are indistinguishable (also for larger $\Wo=20$, see figure~\ref{fig:OPF}a--c). 
	
\section{Conclusion}\label{sec:conclusion}

We showed that the classic lift-up mechanism produces the largest transient growth in pulsatile flow of low amplitude, $A\lesssim0.4$. This is in agreement with recent experiments of \citet{Xu17} and direct numerical simulations of \citet{Xu18}, exhibiting turbulent puff and slugs as in steady pipe flow. At higher amplitudes, helical disturbances begin to dominate in a band of intermediate {Womersley} number $4\lesssim\Wo\lesssim18$ which progressively widens toward larger $\Wo$ as the oscillatory Reynolds number $\Rey_o$ is increased. For the specific case of oscillatory flow ($\A\rightarrow\infty$), our results are in qualitative agreement with the transient growth analysis of \citet{Biau16} for oscillatory Stokes flows over a flat plate. A key difference between oscillatory pipe flow and oscillatory flow over a flat plate is that the former is not solely governed by $\Rey_\delta$, but also by $\Wo$. In particular, the thickness of the Stokes layer in oscillatory pipe flow scales as $\Wo^{-1}$ and at sufficiently low $\Wo$ it fills the pipe. Hence, oscillatory pipe flow is only exactly similar to oscillatory flow over a flat plate in the limit $\Wo\rightarrow \infty$, where curvature effects become negligible \citep{Thomas12}. This convergence can be observed in figure~\ref{fig:OPF}(c); axisymmetric disturbances dominate for $\Rey_\delta\gtrsim200$ at $\Wo=20$, $\Rey_\delta\gtrsim400$ at $\Wo=15$, whereas for $\Wo=10$ helical disturbances dominate even up to  $\Rey_\delta\approx 800$. Note also that the axial wavenumber of the optimal disturbance scaled with $\delta^{-1}$ is close to $0.4$ in figure~\ref{fig:OPF}, which is in excellent agreement with figure~2 of \citet{Biau16}.  

We here considered linear transient growth of disturbances, but transition to turbulence can only be completed with nonlinear effects. \citet{xu2020} showed that initializing direct numerical simulations with the linear optimal helical disturbance can trigger turbulent flow patterns as those observed in their experiments. In experiments of oscillatory pipe flow, the transition threshold is independent of $\Wo$ \citep[with different critical numbers $280 \lesssim \Rey_\delta \lesssim 550$ depending on the setup, see][]{Sergeev66,Merkli75,Hino76,Eckmann91,zhao1996}, which is consistent with our observation that the maximum energy gain depends only on $\Rey_\delta$. {We found large energy growth already at moderately low $\Rey_o = \mathcal{O}(1000)$, which suggests that non-modal disturbance energy growth may be at the root of disturbed flow patterns observed in physiology. Testing this hypothesis requires detailed direct numerical simulation of flows at physiologically relevant parameter regimes and the consideration of non-harmonic waveforms of the flow rate. We anticipate that the steep deceleration phase in aortic flow may be even more prone to instability than in the case of harmonic pulsation.}

\section*{Acknowledgment}
This work was supported by the Deutsche Forschungsgemeinschaft (DFG) in the framework of the research
unit FOR 2688 `Instabilities, Bifurcations and Migration in Pulsatile Flows' under grant AV 120/6-1. Baofang Song acknowledges the financial support from the National Natural Science Foundation of China under grant number 91852105.

\bibliographystyle{jfm}

\bibliography{PPF_linear}

\end{document}